\documentclass[a4paper,11pt]{amsart}
\usepackage{graphicx}
\usepackage{amssymb}
\begin{document}
\pdfoutput=1

\hyphenation{gra-vi-ta-tio-nal re-la-ti-vi-ty Gaus-sian
re-fe-ren-ce re-la-ti-ve gra-vi-ta-tion Schwarz-schild
ac-cor-dingly gra-vi-ta-tio-nal-ly re-la-ti-vi-stic pro-du-cing
de-ri-va-ti-ve ge-ne-ral ex-pli-citly des-cri-bed ma-the-ma-ti-cal
de-si-gnan-do-si coe-ren-za pro-blem gra-vi-ta-ting geo-de-sic
per-ga-mon cos-mo-lo-gi-cal gra-vity cor-res-pon-ding
de-fi-ni-tion phy-si-ka-li-schen ma-the-ma-ti-sches ge-ra-de
Sze-keres con-si-de-red tra-vel-ling ma-ni-fold re-fe-ren-ces
geo-me-tri-cal in-su-pe-rable sup-po-sedly at-tri-bu-table
Bild-raum in-fi-ni-tely counter-ba-lan-ces iso-tro-pi-cally
ap-proxi-mate}

\title[``Information Paradox'' and Schwarzschildian geodesics]
{{\bf ``Information Paradox'' \\and Schwarzschildian geodesics}}

\author[Angelo Loinger]{Angelo Loinger}
\address{A.L. -- Dipartimento di Fisica, Universit\`a di Milano, Via
Celoria, 16 - 20133 Milano (Italy)}
\author[Tiziana Marsico]{Tiziana Marsico}
\address{T.M. -- Liceo Classico ``G. Berchet'', Via della Commenda, 26 - 20122 Milano (Italy)}
\email{angelo.loinger@mi.infn.it} \email{martiz64@libero.it}

\vskip0.50cm

\begin{abstract}
We show that the ``Information Paradox'' follows from
inappropriate considerations on the geodesics of a
Schwarzschildian manifold created by a gravitating point-mass. In
particular, we demonstrate that the geometric differential
equation which gives the radial coordinate as a function of the
angular coordinate of the geodesics does not represent fully all
the consequences following from the metric tensor. We remark that:
\emph{i}) it does not yield the conditions characterizing the
circular orbits; (this fact has been ignored in the previous
literature); \emph{ii}) it ``neglects'' the space region in which
the radial coordinate is minor or equal to twice the mass of the
gravitating point (in suitable units of measure).
\end{abstract}

\maketitle

\vskip1.20cm \noindent \small \textbf{Summary} -- \textbf{1}.
Introduction and r\'esum\'e of the main theses. -- \textbf{1bis}.
On Kundt's physical explanations of the observational data about
the believed BHs. --  \textbf{2}, \textbf{3}, \textbf{3bis}. A
precise treatment \emph{\`a la} Hilbert of the geodesics of a
Schwarzaschild's manifold. -- \textbf{4}. Physical meaning of the
$t$-parametrization of the mentioned geodesics. -- \textbf{5}.
Independently of any specific instance, the formal structure of GR
excludes the existence of any ``Information Paradox''. --
\textbf{6}. \emph{Legenda} regarding De Jans' diagrams of
geodesics at the end of the present paper. -- \emph{Appendix A}:
Computative verifications of the inadequacy of geometric eq.
(\ref{eq:seven}) in  the treatment of the circular orbits. --
\emph{Appendix B}: On binaries composed of two mass-points
according approximate calculations of Numerical Relativity founded
on a $(3+1)$-decomposition of Einstein field equations. --
\par Diagrams of Schwarzschildian geodesics. --

\normalsize

\vskip1.20cm \noindent \textbf{1.} -- A glance over the recent
literature about the Einsteinian gravitating point-mass shows that
the subject is still interesting \cite{1}. The diffuse
interpretation of Schwarzschild's solution which has given origin
to the notion of black hole (BH) is unfounded, see \cite{2}. A
unpleasant consequence of this notion is the belief in the
so-called ``Information Paradox'', according to which GR would be
in contradiction with the time reversibility. Indeed, it has been
affirmed that in the instance of Schwarzschild's manifold created
by a gravitating mass-point the test-particles and the light-rays
go beyond the space surface $R(r)=2m$ -- where $R(r)$ is the
radial coordinate \cite{3} and $m$ is the mass of the gravitating
point $(c=G=1)$ -- and disappear from the ``external'' world for
ever, with an irreversible process. We shall show in this Note
that if one takes into account \textbf{all} the assumptions which
characterize the deduction of the geometric differential equation
of the geodesic trajectories $[R(r)=$ a function of $\varphi,
(0\leq \varphi \leq 2\pi)]$ of test-particles and light-rays, one
obtains a confirmation of the \emph{dynamical} results by Droste
\cite{4} and De Jans \cite{5}: the geodesics that arrive on the
surface $R(r)=2m$ find here their end: \emph{the ``Information
Paradox'' does not exist}.

\par Of course, this conclusion is implicitly contained in
\cite{2}, but we think useful to give an explicit and detailed
proof of the erroneousness of a widespread belief.

\vskip1.20cm \noindent \textbf{1bis.} -- Astrophysics is an
observational and experimental science. All phenomena that the
current ``$\delta \acute{o} \xi \alpha$'' ascribes to BHs can be
actually explained in quite physical ways -- see, \emph{e.g.},
Kundt \cite{6}. According to this Author: \emph{i}) the believed
stellar-mass BHs are neutron stars inside accretion disks;
\emph{ii}) the central engine of an Active Galactic Nucleus (AGN)
is a nuclear-burning disk.

\vskip1.20cm \noindent \textbf{2.} -- Schwarzschild's manifold of
a gravitating point-mass $m$ is characterized by the following
$\textrm{d}s^{2}$:

\begin{equation} \label{eq:one}
\textrm{d}s^{2} = \frac{R(r)}{R(r)-\alpha} \, [\textrm{d}R(r)]^{2}
+ [R(r)]^{2} (\textrm{d}\vartheta^{2} + \sin^{2}\vartheta \,
\textrm{d}\varphi^{2})- \frac{R(r)-\alpha}{R(r)} \,
\textrm{d}t^{2} \, ; \quad (c=G=1) \, ,
\end{equation}

where $\alpha \equiv 2m$, and $R(r)$ is a regular function of $r$
such that the $\textrm{d}s^{2}$ becomes Minkowskian if $r
\rightarrow \infty$. (In the standard solution $R(r) \equiv r$, in
the original Schwarzschild's solution
$R(r)=(r^{3}+\alpha^{3})^{1/3}$, in Brillouin's solution $R(r)
\equiv r+\alpha$, \emph{etc.}). The geodesics of test-particles
and light-rays are plane trajectories and obey the following
equations -- see \cite{7}, eqs. (41)$\div$(44):

\begin{equation} \label{eq:two}
\frac{R}{R-\alpha} \, \left(
\frac{\textrm{d}R}{\textrm{d}p}\right)^{2} + R^{2} \, \left(
\frac{\textrm{d}\varphi}{\textrm{d}p}\right)^{2} -
\frac{R-\alpha}{R} \, \left(
\frac{\textrm{d}t}{\textrm{d}p}\right)^{2} = A \, (\leq 0) \quad;
\end{equation}

\begin{equation} \label{eq:three}
R^{2} \, \left( \frac{\textrm{d}\varphi}{\textrm{d}p}\right) = B
\quad;
\end{equation}

\begin{equation} \label{eq:four}
\frac{R-\alpha}{R} \, \frac{\textrm{d}t}{\textrm{d}p} = C \quad;
\end{equation}

\begin{equation} \label{eq:five}
\frac{\textrm{d}}{\textrm{d}p} \left( \frac{2R}{R-\alpha} \,
\frac{\textrm{d}R}{\textrm{d}p} \right) +
\frac{\alpha}{(R-\alpha)^{2}} \, \left(
\frac{\textrm{d}R}{\textrm{d}p}\right)^{2} - 2R \,
\left(\frac{\textrm{d}\varphi}{\textrm{d}p}\right)^{2} +
\frac{\alpha}{R^{2}} \, \left(
\frac{\textrm{d}t}{\textrm{d}p}\right)^{2} = 0 \quad.
\end{equation}

$A, B, C$ are integration constants (which respect to the affine
parameter $p$); $A$ is zero for the light-rays, negative for the
test particles; we can put $C=1$, by virtue of the arbitrariness
of $p$. The Lagrangean eq. (\ref{eq:five}) for $R$ is connected
with eqs. (\ref{eq:two}), (\ref{eq:three}), (\ref{eq:four});
indeed, we have the identity:

\begin{equation} \label{eq:six}
\frac{\textrm{d}[2]}{\textrm{d}p} - 2 \,
\frac{\textrm{d}\varphi}{\textrm{d}p} \,
\frac{\textrm{d}[3]}{\textrm{d}p}  + 2 \,
\frac{\textrm{d}t}{\textrm{d}p} \,
\frac{\textrm{d}[4]}{\textrm{d}p} =
\frac{\textrm{d}[R]}{\textrm{d}p} \, [5] \quad,
\end{equation}

where the brackets denote the left sides of eqs. (\ref{eq:two}),
(\ref{eq:three}), (\ref{eq:four}), (\ref{eq:five}). The
elimination of $\textrm{d}p$ and $\textrm{d}t$ from
(\ref{eq:two}), (\ref{eq:three}), (\ref{eq:four}) gives the
\emph{geometric} differential equation of the geodesics:

\begin{equation} \label{eq:seven}
\left( \frac{\textrm{d}\varrho}{\textrm{d}\varphi}\right)^{2} =
\frac{1+A}{B^{2}} - \frac{A \, \alpha}{B^{2}} \, \varrho -
\varrho^{2} + \alpha \varrho^{3} \quad,
\end{equation}

where $\varrho:= 1/R$. Since for the \emph{circular} orbits
$\textrm{d}R/ \textrm{d}p=0$, in this case identity (\ref{eq:six})
is \emph{not} a consequence of (\ref{eq:two}), (\ref{eq:three}),
(\ref{eq:four}). Consequently, as it is easy to verify, eq.
(\ref{eq:seven}) \emph{does not give the correct restrictions on
the above orbits} (see Appendix A).

\vskip1.20cm \noindent \textbf{3.} -- Of course, eq.
(\ref{eq:two}) implies $R>\alpha$, \emph{but eq.} (\ref{eq:seven})
\emph{``neglects'' this condition}. The  substitutions $(R-\alpha)
\rightleftarrows -t$ give for $R \leq \alpha$ a \emph{non}-static
metric for which the temporal and the radial coordinates
interchange their roles; in particular, eq. (\ref{eq:seven})
becomes:

\begin{equation} \label{eq:sevenprime}
\left( \frac{\textrm{d}[1/(\alpha-t)]}{\textrm{d}\varphi}
\right)^{2} = \frac{1+A}{B^{2}} - \frac{A \, \alpha}{B^{2}} \,
\frac{1}{(\alpha-t)} - \frac{1}{(\alpha-t)^{2}} +
\frac{\alpha}{(\alpha-t)^{3}} \tag{7$'$}
 \quad.
\end{equation}

This is \emph{not}, however, a significant result, because the
geodesic parametrization with $p$ -- or with the proper time $s$
-- and the parametrization with $\varphi$ of eq. (\ref{eq:seven})
-- would give a geodesic surpassing of $R=\alpha$ \emph{with the
original coordinates} $R$ and $t$.

\par For the circular orbits eq. (\ref{eq:five}) gives:

\begin{equation} \label{eq:fiveprime}
-2R \left(\frac{\textrm{d}\varphi}{\textrm{d}p}\right)^{2}
 + \frac{\alpha}{R^{2}} \, \left(\frac{\textrm{d}t}{\textrm{d}p}\right)^{2}= 0 \tag{5$'$}\quad;
\end{equation}

from which the circular velocity $v$:

\begin{equation} \label{eq:fivesecond} v^{2} =  \left(R \, \frac{\textrm{d}\varphi}{\textrm{d}p}\right)^{2} = \frac{\alpha}{2R} \tag{5$''$}
\quad.
\end{equation}

For the test-particle geodesics, we have from eq. (\ref{eq:two})
-- with $A<0$ -- and eq. (\ref{eq:fiveprime}) that

\begin{equation} \label{eq:eight}
R > \frac{3}{2} \, \alpha \quad ,
\end{equation}

\begin{equation} \label{eq:eightprime}
v < \frac{1}{\sqrt{3}}\tag{8$'$} \quad .
\end{equation}

And for the light-rays $(A=0)$:

\begin{equation} \label{eq:nine}
R = \frac{3}{2} \, \alpha \quad ;
\end{equation}

\begin{equation} \label{eq:nineprime}
v = \frac{1}{\sqrt{3}} \tag{9$'$} \quad ,
\end{equation}

The restrictions (\ref{eq:eight}), (\ref{eq:eightprime}) and
(\ref{eq:nine}), (\ref{eq:nineprime}) are not deducible from the
geometric equation (\ref{eq:seven}).

\vskip1.20cm \noindent \textbf{3bis.} -- The metric generated in
the spòace domain $R(r)\leq \alpha$ by the substitutions
$R(r)-\alpha \rightleftarrows -t$ is a non-static metric for a
static problem. From the standpoint of the differential geometry,
this fact does not represent a difficulty. But from the physical
point of view, things stand otherwise, because the non-static
character implies clearly the existence of \emph{transport}
forces, that are extraneous to our problem. This means that the
above metric is only a formal trick, which cannot give a physical
significance to space domain $R(r)\leq \alpha$, that in reality
does \emph{not} belong to Schwarzschild's manifold. Remark that
for the radial coordinates by Schwarzschild and by Brillouin (cf.
sect. \textbf{2}) this space domain is reduced to a singular
point.

\par We emphasize finally that also the well-known metric of
Kruskal and Szekeres is non-static, and therefore introduces
transport forces in a static problem.

 \vskip1.20cm \noindent \textbf{4.} -- As it was emphasized by
 von Laue \cite{8}, in Schwarzschild's manifold of a gravitating
 material point the \emph{Systemzeit} $t$ has a clear physical
 meaning, as it is specially attested by the red-shift of the
 spectral lines.

 \par Now, with the $t$-parametrization of the dynamical evolution
 -- which is privileged by Droste \cite{4} --, we have that the
 velocities and the accelerations of the geodesics at $R=\alpha$
 are equal to \emph{zero}: Hilbertian repulsion by the event
 horizon $R(r)= \alpha$. \cite{7}.

\vskip1.20cm \noindent \textbf{5.} -- Back to the ``Information
Paradox''. We could affirm \emph{a priori}, \emph{i.e.} without
the detailed examination of the geodesics in a Schwarzschildian
manifold of a gravitating point-mass, that it cannot have a real
existence in a theory as the GR, that has been devised in a manner
which is independent of the directions of the spacetime
coordinates, and in particular independent of the direction of the
temporal coordinate.

\par Any contradiction to this fact must be ascribed to an
erroneous interpretation of a given aspect of the formalism.

\vskip1.20cm \noindent \textbf{6.} -- At the end of paper
$[5$\emph{b})$]$, De Jans emphasizes that the solutions of eq.
(\ref{eq:seven}) can be divided into four categories, that he
illustrates with some diagrams. For each figure he gives the
values of $\sigma \equiv -\alpha^{2} \, A/B^{2}$ and $\tau \equiv
\alpha^{2} / \beta^{2}$, where $A, B$ are the constants of our
eqs. (\ref{eq:two}) and (\ref{eq:three}). For the radial and
circular geodesics we have no figure. It is remarkable that
\emph{for no geodesic there is the surpassing of the space
surface} $R=\alpha$.

\par \emph{Categorie A}. -- Orbits with pericentre: periodic orbits (Figs. 1a and 1b); limiting
orbits; open orbits (Figs. 2a and 2b).

\par \emph{Categorie B}. -- Finite orbits, with apocentre and
without pericentre (Figs. 3a and 3b).

\par \emph{Categorie C}. -- Finite orbits, with apocentre and
without pericentre (Fig. 4); infinite orbits without apsides,
without asymptote; infinite orbits without apsides, with an
asymptote (Figs. 5a and 5b); radial orbits.

\par \emph{Categorie D}. -- Transition orbits between categories B and C (cf. Figs. 3); orbits with apocentre
and internal asymptotic circle (Fig. 6); orbits without apocentre,
with internal asymptotic circle, without asymptote; orbits without
apocentre, with internal asymptotic circle, with an asymptote
(Fig. 7); orbits without pericentre and with internal asymptotic
circle (Fig. 8).

\par The subdivision into the above categories depends on the
discriminant $\Delta$ of Weierstra\ss{}' elliptic function
$\mathcal{P}$ which gives the general solution of eq.
(\ref{eq:seven}):

\begin{equation} \label{eq:ten}
\frac{\alpha}{4R} = \mathcal{P} \, (\varphi+K)+ \frac{1}{12} \quad
,
\end{equation}

where $K$ is a constant of integration; $\Delta$ is given by the
following equality:

\begin{equation} \label{eq:tenprime}
\Delta := g_{2}^{3} -  g_{3}^{2} \tag{10$'$} \quad ,
\end{equation}

where:

\begin{equation} \label{eq:eleven}
g_{2} := \frac{1}{12} + \frac{\alpha^{2}}{4} \, \frac{A}{B^{2}}
\quad .
\end{equation}

\begin{equation} \label{eq:elevenprime}
g_{3} :=  \frac{1}{216} \left( 1 - 9 \, \alpha^{2} \frac{A}{B^{2}}
- \frac{27 \alpha^{2}}{2} \, \frac{1}{B^{2}} \right) \tag{11$'$}
\quad .
\end{equation}

We have: $\Delta>0$ for \emph{Categorie A}, \emph{Categorie B};
$\Delta<0$ for \emph{Categorie C}; $\Delta=0$ for \emph{Categorie
D} and for the circular orbits. --

\par See the diagrams of $[$5$b$)$]$ at the end of the present
paper; \emph{they are referred to the standard radial coordinate}
$R(r) \equiv r$.

\vskip2.00cm
\begin{center}
\noindent \small \emph{\textbf{APPENDIX A}}
\end{center} \normalsize

\vskip0.40cm \noindent To show the inadequacy of eq.
(\ref{eq:seven}) in the treatment of the circular geodesics, it is
sufficient to consider the instance of the light-rays, for which
$A=0$. Eq. (\ref{eq:seven}) becomes:

\begin{equation} \label{eq:A1}
\left( \frac{\textrm{d}(1/R)}{\textrm{d}\varphi} \right)^{2} =
\frac{1}{B^{2}} - \frac{1}{R^{2}} + \frac{\alpha }{R^{3}} \tag{A1}
\quad .
\end{equation}

For a circular orbit we must have:

\begin{equation} \label{eq:A2}
\frac{1}{B^{2}} = \frac{1}{R^{2}} - \frac{\alpha }{R^{3}} \tag{A2}
\quad .
\end{equation}

Let us put:

\begin{equation} \label{eq:A3}
R = k \, \frac{\alpha}{2} \quad, \quad \textrm{with} \quad k\geq 3
\quad; \tag{A3} \quad .
\end{equation}

then:

\begin{equation} \label{eq:A4}
\frac{1}{B^{2}} = \frac{1}{k^{3}\,\alpha^{2}} \, (4k-8)
> 0  \tag{A4} \quad ;
\end{equation}

in particular, for $k=3$:

\begin{equation} \label{eq:A5}
\frac{1}{B^{2}} = \frac{1}{27\alpha^{2}}
 \tag{A5} \quad ,
\end{equation}

which is the \emph{unique} value prescribed by the dynamical
solution. We see, however, that the geometric eq. (\ref{eq:seven})
allows all the trajectories for which $k>3$.

\par If $k= (3-\eta)$, with $\eta>0$ and $\leq 3$, the
right-hand side of (\ref{eq:A4}) becomes
$\frac{4}{(3-\eta)^{3}\alpha^{2}} \, (1-\eta)$. We see that for
$\eta\geq 1$ no circular orbit is possible.

\vskip2.00cm
\begin{center}
\noindent \small \emph{\textbf{APPENDIX B}}
\end{center} \normalsize

\vskip0.40cm \noindent The Numerical Relativity investigates, in
particular, existence and behaviour of binaries composed of two
mass-points with approximate calculations that hardly can have an
exact counterpart \cite{9}. \emph{A fortiori}, this consideration
holds for the three believed supermassive BHs residing in a quasar
triplet \cite{10}. Clearly, also for these instances no
``Information Paradox'' exists.

\par The numerical computations make use of $(3+1)$decompositions
of Einstein field equations. Now, a $(3+1)$-decompositions is
\emph{not} fully equivalent to Einstein gravitational theory,
which considers also reference frames corresponding to metrics not
belonging to the class of metrics characterized by any
$(3+1)$-decomposition. The $(3+1)$-decompositions  are an obvious
generalization of the Gaussian frames, which were devised by
Hilbert in 1916 \cite{7}. In general, the metrics of GR must only
satisfy the well-known conditions that $g_{44}$ be negative and
the quadratic form with the coefficients $g_{\alpha\beta},
(\alpha, \beta = 1, 2, 3)$, be positive-definite.

\par Finally, it is contrary to the spirit of GR to ascribe a
conceptual importance to any spacetime ``foliation''. From a
purely geometric standpoint, a given ``foliation'' has the same
value as a given Gaussian frame.

\newpage \vskip3.00cm
\begin{figure}[!ht]
\begin{center}
\includegraphics[width=0.8\textwidth]{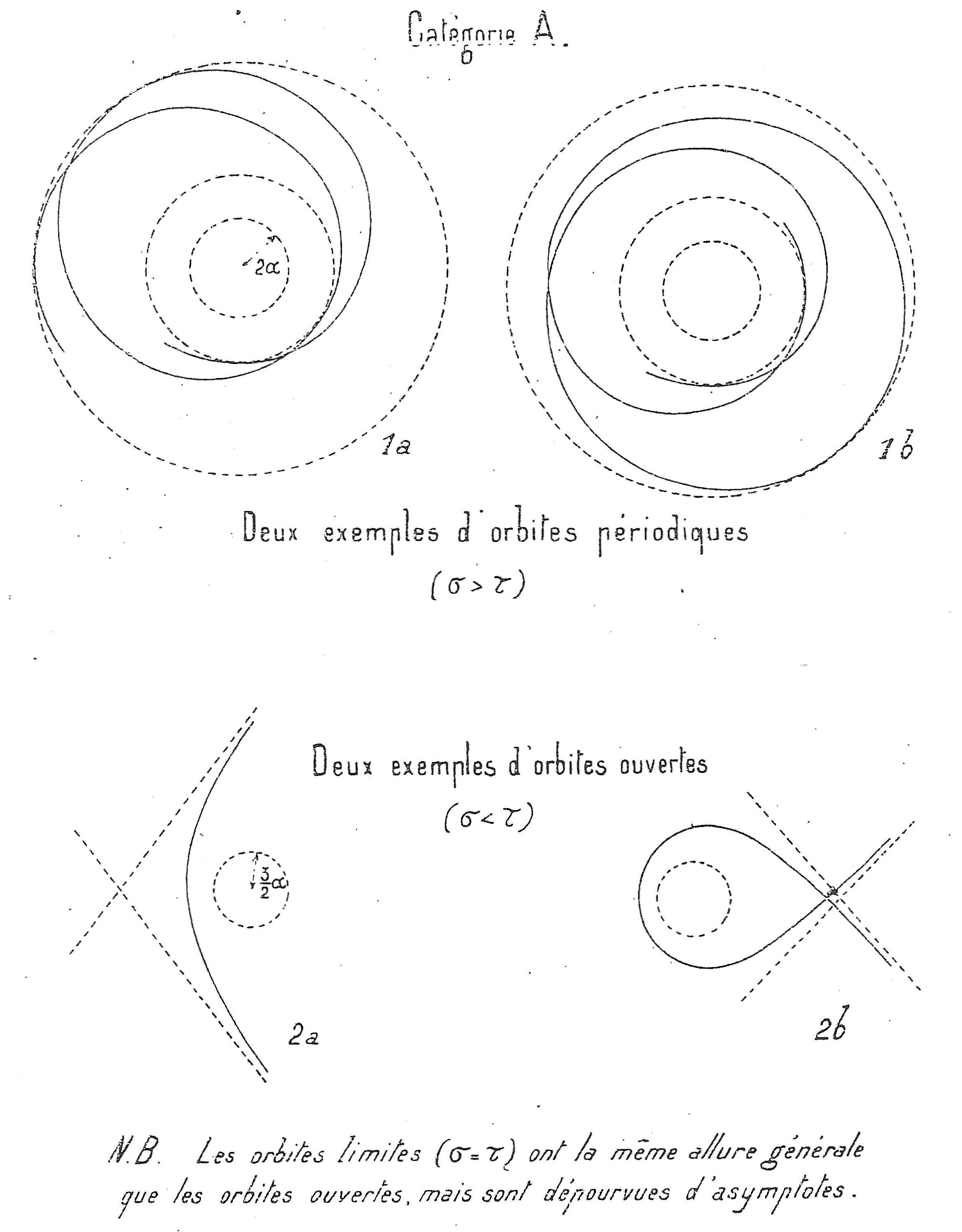}
\end{center}
\end{figure}
\normalsize

\newpage
\begin{figure}[!ht]
\begin{center}
\includegraphics[width=0.8\textwidth]{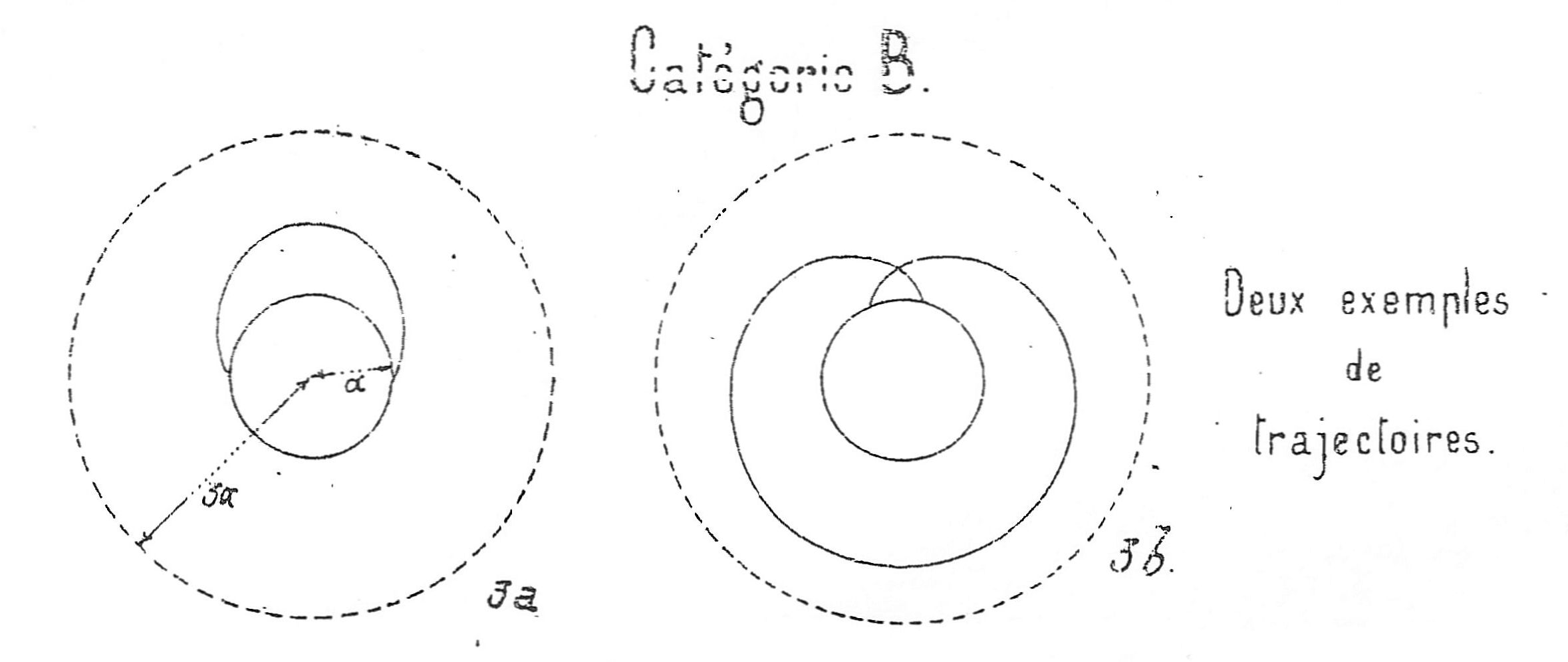}
\end{center}
\end{figure}
\normalsize

\vskip3.00cm

\begin{figure}[!ht]
\begin{center}
\includegraphics[width=0.8\textwidth]{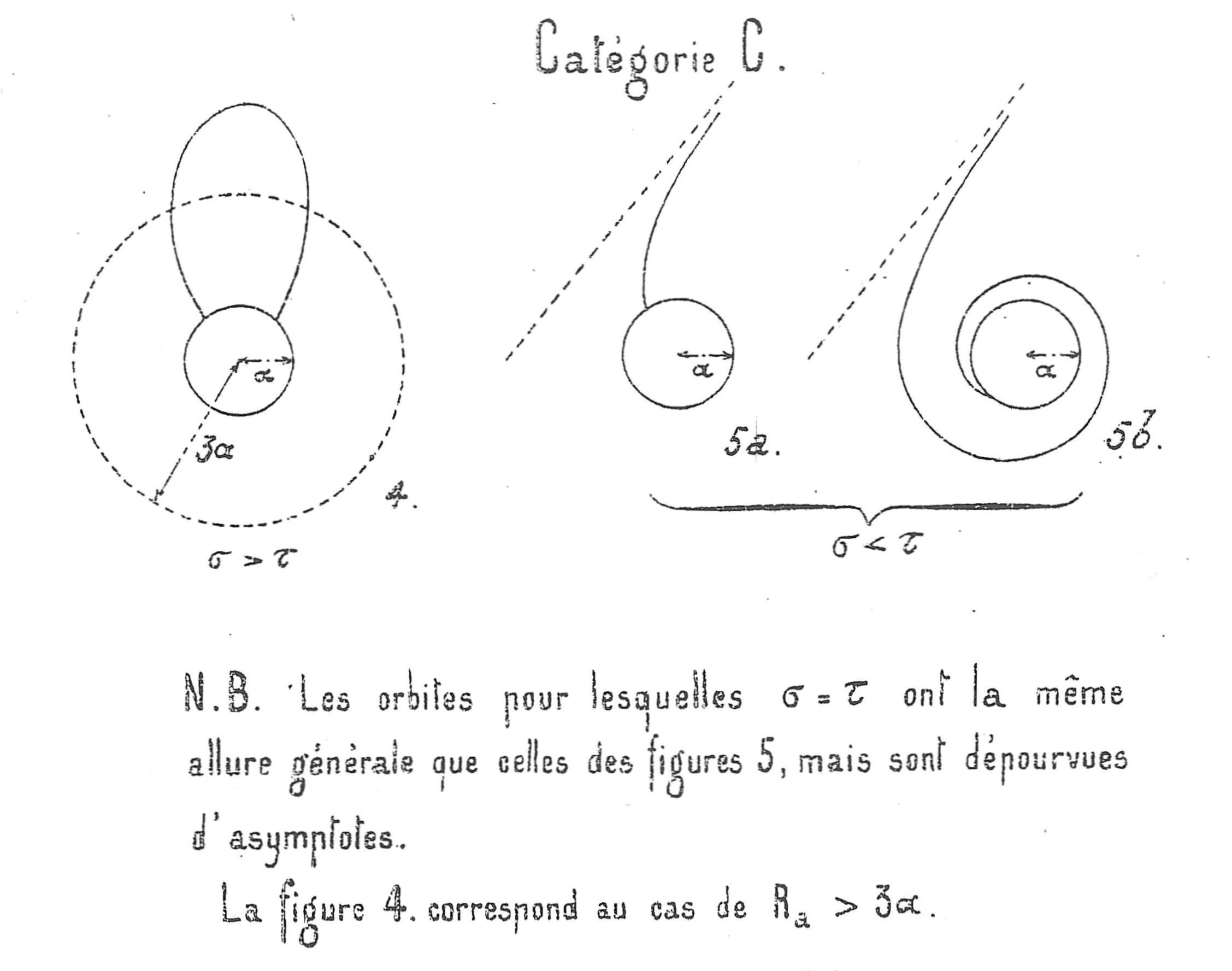}
\end{center}
\end{figure}

\newpage
\begin{figure}[!ht]
\begin{center}
\includegraphics[width=0.8\textwidth]{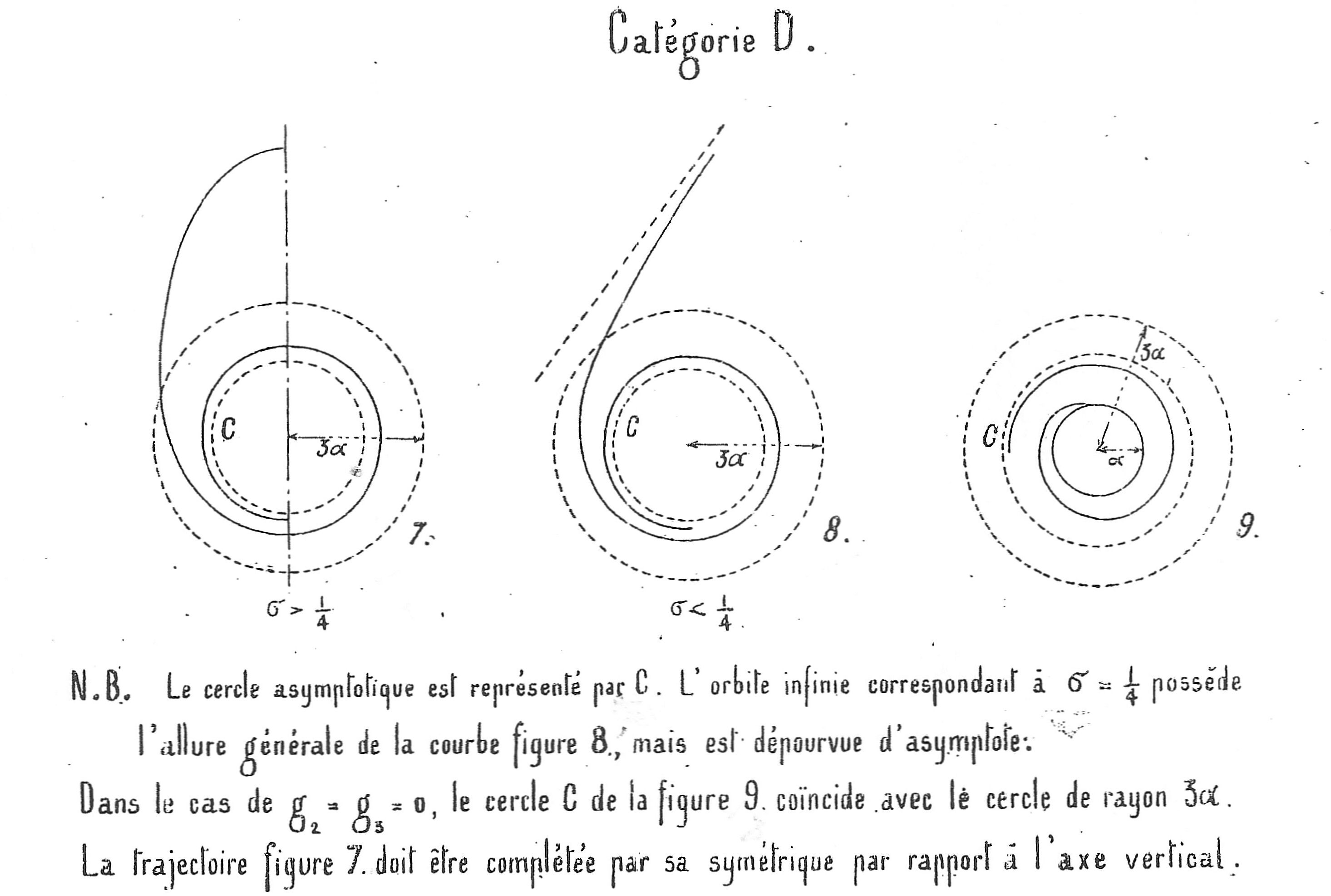}
\end{center}
\end{figure}
\small Par suite d'une erreur, le nombre 6 a \'et\'e omis dans la
num\'erotation des figures $[5$\emph{b})$]$, p.91.-- 

\end{document}